\documentclass[preprint2]{aastex}
\usepackage{graphicx}
\usepackage{color}

\begin{document}


\title{Pulsars as natural launchers of interstellar probes}

\author{Z.N. Osmanov}
\affil{School of Physics, Free University of Tbilisi, 0183, Tbilisi,
Georgia}

\affil{E. Kharadze Georgian National Astrophysical Observatory, Abastumani 0301, Georgia}

\begin{abstract}
In this study, we investigate the feasibility of accelerating small probes up to relativistic energies via magnetocentrifugal acceleration by pulsars. Specifically, we consider a broad range of pulsars and examine probes with radii ranging from $10$ to $1000$ nm. We show that the probes may attain Lorentz factors of up to approximately $25$. Furthermore, we investigate the detectability of such probe swarms and demonstrate that, if a highly advanced civilization deploys probes with a total mass of approximately one billion tonnes for acceleration, their collective synchrotron emission could be detectable by the FAST radio telescope at a distance of 1 kpc. The study also discusses the potential resemblance between the emission from such probe swarms and fast radio bursts (FRBs), and outlines directions for future research.

\end{abstract}


\section{Introduction}
The primary obstacle to interstellar travel is energy. The kinetic energy required to launch a probe increases rapidly with velocity as relativistic effects become significant, representing a substantial resource cost even for a technologically advanced civilization. To mitigate this problem, the use of external energy sources has been proposed, including laser- or microwave-driven light sails \citep{forward,lubin} and acceleration through natural gravitational systems \citep{dyson}.

The concept of self-replicating probes, based on von Neumann’s theory of self-reproducing automata \citep{vneum}, further reduces the initial resource requirements. Such probes (von Neumann probes, or VNPs) could manufacture copies of themselves using local resources in each star system, potentially spreading throughout the Galaxy within a relatively short period \citep{freitas}. In recent years, both the relevance of this concept as a new avenue for SETI research and its technological feasibility have been discussed \citep{matloff, borque,probe3,probe2,probe1}. According to the arguments advanced by \cite{hart}, the apparent absence of such probes from astronomical observations sharpens the Fermi paradox. Investigating the physical mechanisms that could enable probe launches is therefore also relevant to the broader discussion of this paradox. This question is particularly important in light of studies \citep{probe3,probe2,probe1} arguing that such VNPs could be extremely small.

In this context, an intriguing question is whether nature provides objects that could be used to accelerate probes without requiring additional investment by a civilization. One potential candidate is a pulsar. Pulsars, rapidly rotating magnetized neutron stars \citep{hewish}, convert rotational kinetic energy into radiation and particle winds \citep{GJ}. The spin-down power of young pulsars can exceed ($10^{38}$ erg/s), while they are exceptionally stable and long-lived energy sources. These properties make pulsars potential natural launch platforms for probes, not only self replicating VNPs. 

It is worth noting that pulsars are highly efficient sources of ultrahigh-energy particles. In particular, magnetocentrifugal effects can accelerate particles to ultrahigh energies up to several PeV \citep{review}. Therefore, endowing the probes with an electric charge could, in principle, enable their acceleration through a similar mechanism.

In this paper, we investigate pulsars as potential natural systems for launching interstellar probes. Although self-replication is beyond the scope of the present study, previous publications have shown that, for VNPs to operate most efficiently, they should have the small masses. We determine the achievable probe velocity/Lorentz factor as a function of its mass and other relevant parameters, examine the limiting factors, and assess the potential observational signatures of this process. 

The remainder of the paper is organized as follows. Sect. 2 describes pulsar energetics and the acceleration model, Sect. 4 presents the results and in Sec. 5 we outline them.

\section[]{Acceleration mechanism}
We begin this section by considering centrifugal acceleration. As is well known, pulsars are neutron stars with extremely strong magnetic fields ($B\gtrsim 10^{12}$ G). Consequently, a charged particle entering such a magnetic field begins to gyrate with a Larmor radius that decreases as the magnetic field strength increases. In particular, for typical probe parameters (we will see this below) $q\simeq 0.02\;StatC$, $m\simeq 10^{-12}\;g$, and the magnetic induction, $B\simeq 10^{12}\; G$, the Gyroradius for a mildly relativistic scenario ($\gamma\simeq 1$) gives $R_g\simeq 2.4\times 10^{-6}$ cm, which is by many orders of magnitude smaller than a typical scale of the pulsar, $10^6$ cm. Here, $q$ and $m$ denote the charge and mass of the probe, respectively, and $\gamma$ represents the relativistic factor of the VNP. The aforementioned result confirms that the VNP follows a magnetic field line which, in turn, rotates at the same angular velocity as the pulsar. Therefore, the probe is expected to undergo efficient magnetocentrifugal acceleration.

For simplicity we consider straight magnetic field lines lying in the equatorial plane. To describe the dynamics of a particle moving along such a field line, we define the Lagrangian of a massive particle
\begin{equation}
\label{L} 
L = -mc\left(1-\frac{\upsilon^2}{c^2}-\frac{\omega^2r^2}{c^2}\right)^{1/2},
\end{equation}
where $r$ is the radial coordinate, $\upsilon = \dot{r}$ represents the corresponding velocity component,  $c$ denotes the speed of light and $\omega$ is the angular velocity of rotation. Since $L$ does not depend on time explicitly, the Hamiltonian $H = \dot{r}\frac{\partial L}{\partial\dot{r}}-L$ is a conserved quantity, and one can obtain an expression of the Lorentz factor \citep{review}
\begin{equation}
\label{gamma} 
\gamma = \frac{\gamma_0}{1-\frac{\omega^2r^2}{c^2}},
\end{equation}
where $\gamma_0$ is an initial relativistic factor of the particle. As is evident from the expression obtained, the Lorentz factor of a centrifugally accelerated particle increases asymptotically as it approaches the light cylinder (LC - a hypothetical area where the linear velocity of rigid rotation exactly equals the speed of light), having the radius $R_{lc} = c/\omega$

However, it is clear that acceleration cannot continue indefinitely and that there must be factors that eventually halt the acceleration process. As discussed above, magnetocentrifugal acceleration is enabled by the particle's average motion along magnetic field lines, which is maintained by the magnetic Lorentz force ${\bf F_L} = q{\bf\upsilon_{\perp}}\times{\bf B}/c$, where $q$ is the probe's charge and $\upsilon_{\perp}$ the velocity component perpendicular to the magnetic field. On the other hand, the rotating field lines exert a perpendicular reaction force on the particles $F = m\omega\left(2\gamma\dot{r}+r\dot{\gamma}\right)$ \citep{rieger}, which increases as they approach the light cylinder. When the reaction force approaches the Lorentz force, the particle can no longer remain tied to the field line, and the acceleration process terminates. This happens when the balance condition is satisfied \citep{rieger}
\begin{equation}
\label{balance} 
m\omega\left(2\gamma\frac{dr}{dt}+r\frac{d\gamma}{dt}\right)\simeq\frac{q}{c}\upsilon_{\perp}B.
\end{equation}
\cite{mach} have shown that due to the magnetocentrifugal acceleration, the radial velocity behaves as $\upsilon = \omega r\left(1-\omega^2r^2/c^2\right)^{1/2}$. By taking this and Eq. (\ref{gamma}) (for $\gamma_0 = 1$) into account one can reduce Eq. (\ref{balance}) for $x = \sqrt{\gamma-1/\gamma}$ 
\begin{equation}
\label{x1} 
x^3+x-A = 0,
\end{equation}
where $A = q\upsilon_{\perp}B/(2m\omega c)$. It should be noted that previous studies considered the ultrarelativistic scenarios ($\gamma>>1$), since they focused on elementary particles. In our case, the VNP masses are many orders of magnitude greater than those of electrons and protons\footnote{In the framework of the mentioned approximation we neglected $1/\gamma$.}. As we will show, the probes can nevertheless be accelerated to relativistic energies, although their Lorentz factors will be of the order of $25$. Of the three roots of Eq. (\ref{x}), only one is real and is therefore of interest to us
\begin{equation}
\label{x} 
x = \frac{2}{\sqrt{3}}sinh\left[\frac{1}{3}arsinh\left(\frac{3\sqrt{3}}{2}A\right)\right],
\end{equation}
which automatically defines the relativistic factor, 
\begin{equation}
\label{gamma1} 
\gamma = x^2+\sqrt{x^4+1}.
\end{equation}
%

\section{Discussion and results}
In this section, we apply the acceleration model to tiny probes and investigate how the results depend on various parameters. \cite{probe2,probe1,probe3}, has investigated the problem of optimizing the size of self-replicating probes, and it has been shown that, depending on the technological level of a civilization and the size of the region of space to be explored, it may be advantageous to use extremely small-size VNPs, ranging in from micrometers to fractions of a millimeter.

For a highly advanced civilization seeking to explore space by deploying self-replicating VNPs, the primary objective would naturally be to achieve the highest possible velocity, which requires minimizing the probe's mass. However, a self-replicating system must possess at least minimal information processing capacity. The diameter of the smallest known cells is generally considered to be approximately $250$nm \citep{size}, whereas virus sizes start at 20 nm. Therefore, in our calculations, we begin by considering spherical probes with a radius of $10$ nm.

It is clear that the probes must be electrically charged, and the higher the charge, the more effective the acceleration. On the other hand, increasing the electric charge increases the tensile stress in the material. Since every material has a finite ultimate tensile strength, the maximum charge that can be sustained by a body is also limited for a given material and size.

Assuming that the maximum stress the probe material can withstand is $\sigma$, the maximum charge of a spherical shell of radius $R$ and thickness $\delta$ can be estimated as follows: the electric pressure, $E^2/(8\pi)$ leads to the repulsion of the two hemispheres with a force, $F=\pi R^2\times E^2/(8\pi)$. This force should be balanced by the force in the cross section of the spherical shell, $2\pi R\sigma_m\delta$
$$q = 4\left(\pi R^3\sigma_m\delta\right)^{1/2}\simeq$$
\begin{equation}
\label{q} 
\simeq 1.5\times 10^{-3}\times \left(\frac{\eta}{0.1}\right)^{1/2}\times\left(\frac{R}{10^{-6}\; cm}\right)^{2}\; StatC,
\end{equation}
Here, we have used $\delta\simeq\eta 2R/3$, where $\eta$ is the filling factor of the probe, defined as the fraction of its total volume occupied by material. We have also adopted $\sigma_m\simeq 1.3\times 10^{12}$ dyn/cm$^2$ corresponding to the assumed strength of graphene. 

For the acceleration effect, the strength of the magnetic field is crucial, which depends on two fundamental pulsar parameters: its rotation period, $P$ and the period derivative, $\dot{P}$ for which the magnetic strength on the neutron star's surface is given by \citep{carroll}
\begin{equation}
\label{B} 
B_{st}\simeq 3.2\times 10^{19}\times\sqrt{P\dot{P}}\; G.
\end{equation}

\begin{figure}
  \centering {\includegraphics[width=8cm]{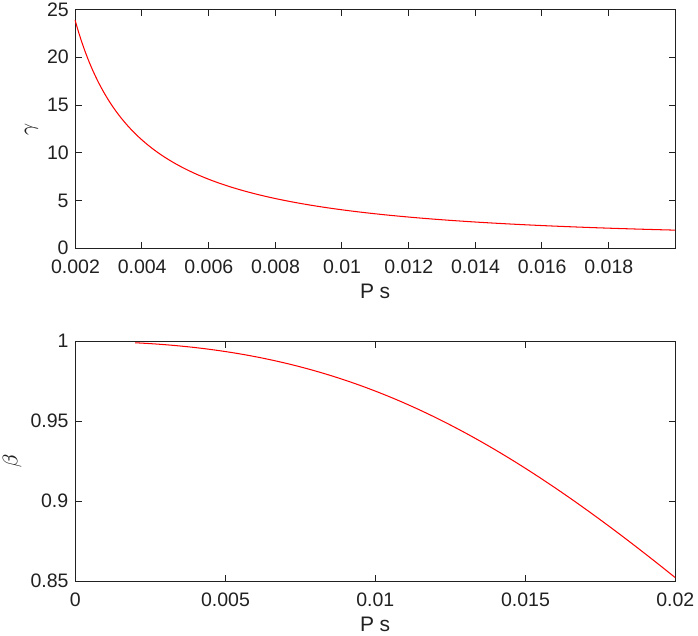}}
 \caption{Plots of $\gamma(P)$ and $\beta(P)$. The set of parameters is: $R = 10\; nm$, $\rho = 2.2\; g/cm^3$, $\dot{P} = 10^{-12}\; ss^{-1}$.}\label{fig1}
\end{figure}

It is estimated that the period derivative of young pulsars, $\dot{P}$ can be of the order of $10^{-12}\; ss^{-1}$, \citep{pdot}, while the fastest known pulsars have spin periods of the order of milliseconds. Assuming that the internal stress in the probes is ten times lower than the maximum allowable stress (thus, $\sigma\simeq 1.3\times 10^{11}\; dyn/cm^2$), after applying Eqs.(\ref{x},\ref{gamma}) and taking into account that the magnetic induction has a dipolar character and on LC is of the order of $B\simeq B_0R_p^3/R_{lc}^3$ we can obtain results, were $R_p\simeq 10^6\; cm$ is a pulsar's radius and $\omega = 2\pi/P$ is the angular velocity of rotation. In Fig. 1 we show the dependence of the Lorentz factor and $\beta = \upsilon/c$ on the rotation period of a newly born pulsars. The set of parameters is: $\upsilon_{\perp} = 10\; km/s$  \footnote{We assume that such velocities can be attained without difficulty, since our own civilization is capable of achieving them, and we consider these civilizations to be far more technologically advanced than ours.} $R = 10\; nm$, $\rho = 2.2\; g/cm^3$, $\dot{P} = 10^{-12}\; ss^{-1}$.

As is evident from the plots, the probe velocities are relativistic, and both the Lorentz factor and the velocity decrease monotonically with increasing pulsar rotation period, as expected.

In our previous studies, we assumed VNP velocities of $1\%$ of the speed of light. However, the present results indicate that the velocities could be very close to the speed of light. On the other hand, probes with a size of $10 nm$ would presumably be able to perform only simple tasks. It is therefore naturally of interest to investigate how the VNP velocities depend on their sizes. The results for this specific case are presented in Fig. 2, which shows the dependence of the Lorentz factor and the $\beta$ factor on the probe size. The set of parameters is the same as in Fig. 1, except for the period of rotation, $P = 0.002\; s$ and $R = [1; 60]\times 10^{-6}\; cm$. As expected, the probe velocity decreases monotonically with increasing probe size. As shown in Fig. 2, the maximum Lorentz factor over the considered range of $R$ can be of the order of $25$.

\begin{figure}
  \centering {\includegraphics[width=8.5cm]{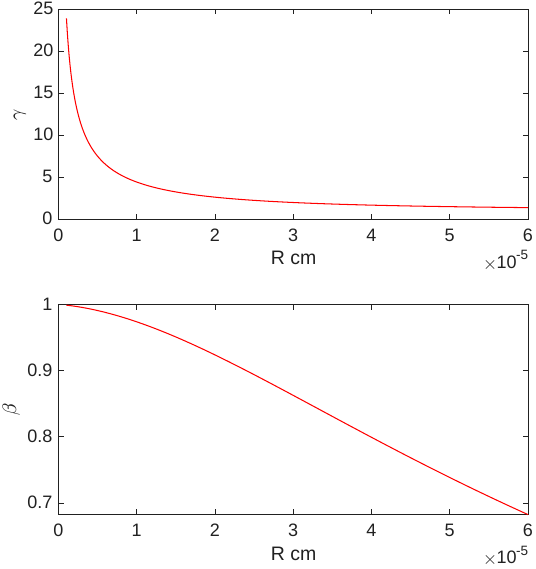}}
 \caption{Plots of $\gamma(R)$ and $\beta(R)$. The set of parameters is the same as in the previous figure, except the probes' sizes and the rotation period, $P = 0.002\; s$ and $R = [1; 60]\times 10^{-6}\; cm$.}\label{fig2}
\end{figure}

\begin{figure}
  \centering {\includegraphics[width=8.5cm]{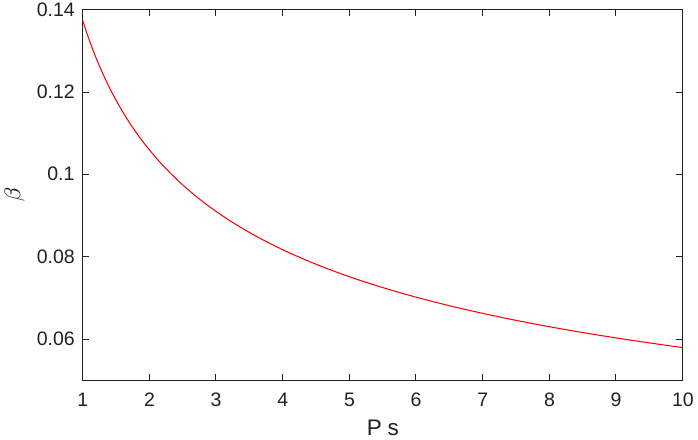}}
 \caption{Plot of the dependence of $\beta$ versus $P$.  The set of parameters is the same as in Fig. 1, except an interval of the rotation period, $P = [1; 10]\; s$ and $\dot{P} = 10^{-15}\; ss^{-1}$}\label{fig3}
\end{figure}

\begin{figure}
  \centering {\includegraphics[width=8.5cm]{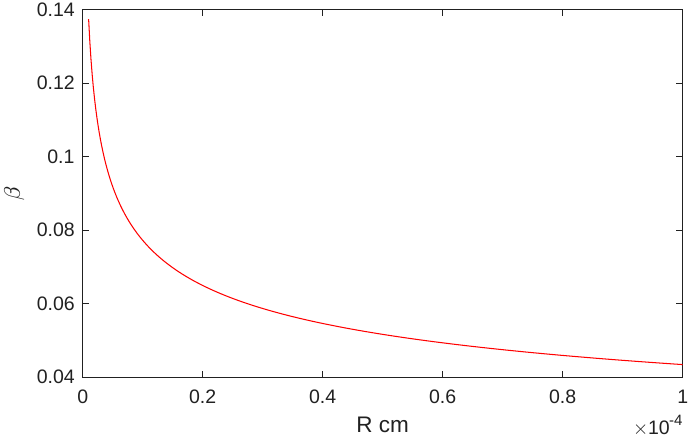}}
 \caption{Plot of $\beta$ versus $R$.  The set of parameters is the same as in the previous figure, except $P = 1 s$ and $R = [1; 100]\times 10^{-6}\; cm$.}\label{fig4}
\end{figure}

In general, in addition to the BBW mechanism, acceleration may also be limited by the radiation reaction force associated with synchrotron emission from charged probes moving in a strong magnetic field. But, as we will demonstrate, its contribution will be negligible. In particular, synchrotron radiation affects the acceleration when the acceleration time scale $t_{acc}\simeq \gamma/\dot{\gamma}\simeq \gamma^{-1/2}P/(4\pi)$ \citep{review} becomes comparable to the radiative cooling timescale $t_{cool}\simeq\gamma mc^2/P_s$, where $P_s\simeq 2q^4B^2\gamma^2/(3m^2c^3)$ denotes the synchrotron emission power \citep{carroll}. Then for the Lorentz factor governed by the synchrotron process, one has
$$\gamma_{syn}\simeq\left(\frac{3\pi m^3c^5}{q^4PB_{lc}}\right)^2\simeq6\times 10^{21}\times\left(\frac{0.1}{\eta}\right)^{4}\times$$
\begin{equation}
\label{gsyn} 
 \times\left(\frac{R}{10\;nm}\times\frac{P}{2\times10^{-3}\; s}\times\frac{10^{12}\; s\; s^{-1}}{\dot{P}}\right)^2.
\end{equation}
Since the Lorentz factor limited by synchrotron radiation is by many orders of magnitude higher than the relativistic factor obtained from the BBW mechanism, this implies that synchrotron losses do not affect the acceleration process.

Magnetocentrifugal acceleration is important not only for millisecond pulsars but is also efficient in the case of ordinary pulsars. To investigate the velocities in this regime, we express the Lorentz factor as $\gamma = 1+\epsilon$ ($\epsilon<<1$), and expand the expression given by Eq. (\ref{x}) in terms of the small parameter, $\epsilon$. Then one can straightforwardly obtain for $\beta$ the following expression
$$\beta\simeq\left(\frac{\sqrt{2}q\beta_{\perp}B_{lc}}{mc\omega}\right)^{1/4}\simeq$$
$$\simeq 0.14\times\left(\frac{1\;s}{P}\right)^{3/8}\times\left(\frac{0.1}{\eta}\right)^{1/8}\times$$
\begin{equation}
\label{beta} 
\times\left(\frac{10\; nm}{R}\times\frac{\upsilon_{\perp}}{10\; km/s}\times\frac{\dot{P}}{10^{-15}\; ss^{-1}}\right)^{1/4},
\end{equation}
where we have assumed a typical value of $\dot{P}$ for relatively slowly rotating pulsars. As is clear from Eq. (\ref{beta}), even for slowly rotating pulsars the velocity is very large.

In Fig. 3 we demonstrate the behavior of $\beta(P)$. The set of parameters is the same as in Fig. 1, except an interval of the rotation period, $P = [1; 10]$ s and $\dot{P} = 10^{-15}\; ss^{-1}$. Although the resulting velocities are not as high as those obtained in the previous cases, it should be borne in mind that ordinary pulsars rotate considerably more slowly. As can be seen, the velocity decreases monotonically with increasing pulsar period, which is consistent with the obtained behavior, $\beta\propto P^{-3/8}$ (see Eq. (\ref{beta})).

In Fig. 4 the behavior of $\beta$ versus the probe sizes is shown, where the set of parameters is the same as in Fig. 3, except $P = 2\; ms$ and $R = [0.1; 10]\times 10^{-5}\; cm$. As expected, for large size VNPs the velocity is smaller. 

In our previous studies, we considered probes traveling at $1\%$ of the speed of light. As we see, even for less energetic scenarios, the velocities will be several times larger. 

However, it is evident that a highly advanced civilization would be interested in harnessing rapidly rotating young pulsars to accelerate probes, since, as we have shown, the probes could reach relativistic energies, with Lorentz factors as high as $25$.

Let us estimate the minimum total mass of probes that a civilization would need to accelerate for them to be detectable at a distance of $d = 1 kpc$. The minimum detectable flux density is given by
\begin{equation}
\label{Smin} 
S_{min} = \frac{2\times T_{sys}\times k_B\times  SNR}{A_{eff}\sqrt{n\Delta\nu t}},
\end{equation}
where $k_B$ denotes the Boltzmann constant and for the FAST telescope, the system noise temperature and effective collecting area are $T_{sys} = 20\; K$  $A_{eff}$ and $7\times 10^4\; cm^2$, respectively. The signal-to-noise ratio, $SNR$, is typically required to be of the order of $5$ and $n = 2$. Here $\Delta\nu$ denotes the bandwidth, and $t$ is the observation time.

Since the so-called L-band of the FAST telescope operates at frequencies of the order of GHz and has a bandwidth of 400 MHz, we need to identify a pulsar whose magnetosphere allows the probes to emit synchrotron radiation in this frequency range. It is straightforward to show that spherical probes with a radius of $100 nm$, moving within the magnetosphere of a pulsar with a period of 2.6 ms, emit in the synchrotron regime with frequency, $\nu\simeq 3\gamma^2qB/(2mc)$ \citep{carroll} precisely in the aforementioned frequency range. Then, for the total luminosity of a swarm of VNPs we will obtain
$$L = 4\pi d^2S_{min}\Delta\nu\simeq$$
\begin{equation}
\label{lumin} 
 \simeq 10^{28}\times\left(\frac{d}{1\;kpc}\right)^2\times\left(\frac{1\;hour}{t}\right)^{1/2}\; erg/s.
\end{equation}
On the other hand, taking into account the power emitted by an individual probe, $P_s\simeq 2q^4B^2/(3m^2c^3)$ \citep{carroll}, one can show that if a highly advanced civilization deploys probes with a total mass of one billion tonnes, the resulting emission would be detectable at a distance of 1 kpc. Since deploying probes with a total mass of one billion tonnes should not pose a significant challenge for a highly advanced civilization, this suggests that such a scenario is entirely plausible.

It is worth noting that such a swarm of probes would undergo acceleration within the pulsar's magnetosphere over a timescale comparable to the pulsar's rotation period. Consequently, the resulting radio emission would be expected to persist for a duration of the same order as the pulsar's rotation period. Thus, the emission would be characterized by short-duration bursts. This behavior bears a striking resemblance to that observed in fast radio bursts - FRB \citep{FRB}. However, a detailed investigation of FRBs requires a separate study, which we plan to undertake in the near future.

The results presented above suggest that a civilization seeking to accelerate probes would favor the use of pulsars with high energy output. Accordingly, observational efforts should primarily focus on high-energy pulsars to search for anomalous behaviors, like FRBs.

\section{Conclusion}
We investigated the feasibility of accelerating small, nano-scale VNPs using pulsars spanning the range from millisecond pulsars to ordinary pulsars.

We have shown that, in the case of millisecond pulsars, the probes can be magnetocentrifugally accelerated to relativistic energies, reaching Lorentz factors of up to $25$.

Although ordinary pulsars rotate much more slowly, we have shown that the probes can still be accelerated to extremely high velocities, reaching at least $4\%$ of the speed of light. This indicates a remarkably efficient acceleration mechanism. We also investigated the acceleration of probes of different sizes, considering radii ranging from $10$ nm to $1000$ nm.

We also estimated the total mass of probes required for their collective emission to be detectable by the FAST telescope at a distance of $1 kpc$. We showed that a total probe mass of approximately one billion tonnes would be sufficient for FAST to detect such a swarm, a mass that should not pose a significant challenge for a highly advanced civilization.

\section*{Acknowledgments}
The research was supported by the Shota Rustaveli National Science Foundation grant (FR-24-1751). 


\begin{thebibliography}{99}

\bibitem[Borque \& Hein(1963)]{borque} Borque, O. \& Hein, A., 2021, Near-term self-replicating probes – a concept design, Acta Astronautica, 187, 546

\bibitem[Cordes \& Chatterjee(2019)]{FRB} Cordes, J.M. \& Chatterjee, S., 2019, Fast Radio Bursts: An Extragalactic Enigma, ARAA, 57, 417

\bibitem[Carroll \& Ostlie(2010)]{carroll} Carroll, Bradley W. \& Ostlie, Dale A., An Introduction to Modern Astrophysics and Cosmology, Cambridge University Press, Cambridge, UK, 2010

\bibitem[Dyson(1963)]{dyson} Dyson, F.J., 1963, Gravitational Machines. In A. G. W. Cameron (ed.), Interstellar Communication: A Collection of Reprints and Original Contributions, 12, 115–120. New York: W. A. Benjamin

\bibitem[Forward(1984)]{forward} Forward, R.L., 1984, Roundtrip interstellar travel using laser-pushed lightsails, Journal of Spacecraft and Rockets, 21, 187

\bibitem[Freitas(1980)]{freitas} Freitas, R.A., 1980, A self-reproducing interstellar probe, JBIS, 33, 251

\bibitem[Goldreich, P. \& Julian(1968)]{GJ} Goldreich, P. \& Julian, W.H., 1969, Pulsar electrodynamics, ApJ, 157, 869

\bibitem[Hart(1975)]{hart} Hart, M.H., 1975, An explanation for the absence of extraterrestrials on Earth, QJRAS, 16, 128.

\bibitem[Hewish et al.(1968)]{hewish} Hewish, A., et al., 1968, Observation of a rapidly pulsating radio source, Nature, 217, 709

\bibitem[Johnston \& Karastergiou(2017)]{pdot} Johnston, S. \& Karastergiou, A., 2017, Pulsar braking and the P–$\dot{P}$ diagram, 467, 3493

\bibitem[Lubin(2016)]{lubin} Lubin, P., 2016, A roadmap to interstellar flight. JBIS, 69, 40

\bibitem[Machabeli \& Rogava(1994)]{mach} Machabeli, G.Z. \& Rogava, A.D., 1994, Centrifugal force: A gedanken experiment, Phys.Rev.A, 50, 98

\bibitem[Matloff(2022)]{matloff} Matloff, G.L., 2022, Von Neumann probes: rationale, propulsion, interstellar transfer timing, IJAb, 21,205

\bibitem[National Research Council(1999)]{size} National Research Council, 1999, Size Limits of Very Small Microorganisms: Proceedings of a Workshop. Washington, DC: National Academy Press

\bibitem[Osmanov(2021a)]{probe3} Osmanov, Z.N., 2021a, Can the China's FAST telescope detect extraterrestrial von-Neumann probes? Serb. Astr. J., 203, 47

\bibitem[Osmanov(2021b)]{review} Osmanov, Z., 2021b, Relativistic Effects of Rotation in $\gamma$-ray Pulsars—Invited Review, Galaxies, 9, 6


\bibitem[Osmanov(2020a)]{probe2} Osmanov, Z.N., 2020a, On a spectral pattern of the Von-Neumann probes, JBIS, 73, 7

\bibitem[Osmanov(2020b)]{probe1} Osmanov, Z.N., 2020b,  On the interstellar Von Neumann micro self-reproducing probes, IJAb, 19,220

\bibitem[Rieger \& Mannheim(2000)]{rieger} Rieger, F.M. \& Mannheim, K., 2000, Particle acceleration by rotating magnetospheres in active galactic nuclei, A\&A, 353, 473

\bibitem[von Neumann(1966)]{vneum} von Neumann, J., 1966, Theory of Self-Reproducing Automata (A. W. Burks, ed.). University of Illinois Press.


\end{thebibliography}
\end{document}